\documentstyle[12pt]{article}

\begin{document}

\begin{titlepage}
\setcounter{page}{1}
\title{\bf On the Staruszkiewicz Modification of the 
Schr\"{o}dinger Equation}
\author{Waldemar Puszkarz\thanks{Electronic address: psi\_bar@yahoo.com}
\\
\small{\it  Department of Physics and Astronomy,}
\\
\small{\it University of South Carolina,}
\\
\small{\it Columbia, SC 29208}}
\date{\small (December 1, 1999)}
\maketitle
\begin{abstract}
%\double
We discuss Staruszkiewicz's nonlinear modification of the  Schr\"{o}dinger 
equation. It is pointed out that the expression for the energy functional for this 
modification is not unique as the field-theoretical definition of energy does not 
coincide with the quantum-mechanical one.  As a result, this modification can 
be formulated in three different ways depending on which physically relevant 
properties one aims to preserve. Some nonstationary one-dimensional solutions 
for suitably chosen potentials, including a KdV soliton, are presented, and the 
question of finding stationary solutions is also discussed. The analysis of 
physical and mathematical features of the modification leads to the conclusion 
that the Staruszkiewicz modification is a very subtle modification of the
fundamental equation of quantum mechanics. 

\vskip 0.5cm
\noindent
%PACS numbers: 11.80.Et, 12.10.Gq, 04.90.$+$e
\end{abstract}
\end{titlepage}

\section{Introduction}

%\double

Some time ago Staruszkiewicz \cite{Sta1}\thinspace put forward what seemed
to be at the time unique\footnote{%
It was demonstrated in \cite{Pusz1} that the Staruszkiewicz modification can
be extended by allowing more Galilean invariant terms of the same property 
in three dimensions.} for three dimensions a way
of modifying the Schr\"{o}dinger equation by adding to its action density a
Galilean invariant term $A_{mod}=\gamma (\Delta S)^{2}/2$, historically
motivated by his theory of the free electromagnetic phase \cite{Sta2, Sta3}.
Here $S$ stands for the phase of the wave function $\Psi =R\exp (iS)$,
originally \thinspace identified with the electromagnetic one. The coupling
constant $\gamma $ was found by some reasoning involving electromagnetic
theory to be equal $1/16\pi ^{2}$ in natural units \cite{Sta2}. This
constant is dimensionless in the system of natural units in three
dimensions, a property that distinguishes the Staruszkiewicz modification
from any other proposed so far. However, neither the identification
mentioned nor any other arguments based on electromagnetic theory are really
necessary as the modification in question can simply be postulated on the
grounds of physical consistency.\footnote{%
In fact, as shown in \cite{Pusz2}\thinspace any fully electromagnetic
extension of the Schr\"{o}dinger equation based on the identification of
these phases runs into conflict with the Galilean invariance of the modified
equation.} In such a case $\gamma $ is an arbitrary constant. It is this
approach that we adopt for our paper. It should be stressed that this is 
a purely field-theoretical approach. As of now the equations proposed by
Staruszkiewicz have not been derived within any phenomenological scheme.
Unfortunately, this is rather disadvantageous for the physical analysis
of the modification as any insight one could derive from such a scheme
cannot be gained here. 

The purpose of this paper is to extend the original brief report \cite{Sta1}%
\thinspace by providing a more elaborate analysis of some aspects of this
interesting modification together with some solutions to it. To this end, in
the next section, we review main points of the modification largely along
the lines of \cite{Sta1} and discuss them in some detail. The following section 
contains a more detailed study of the modification, including a comment on 
the variational derivation of the equations of motion, a discussion of the 
problem of separability of noncorrelated subsystems, remarks on its semiclassical 
approximation and the Ehrenfest relations, as well as other comments on 
its characteristic properties. This section also addresses the problem of 
the energy definition, which for nonhomogeneous nonlinear modifications as 
the one in question is not unique \cite{Pusz3}. Depending on how one chooses 
to approach this issue one can have two different formulations of the
modification. Yet another formulation can be adopted to cirmcumvent the
problem of nonseparability in the approach of Bia\l ynicki-Birula and
Mycielski \cite{Bial}. In another section, concerned with solutions, guided
by earlier observations, we find some characteristic and physically
important solution, a coherent state. It is in this section that we also
demonstrate how to reduce the continuity equation to the form of the
Korteweg-de Vries equation and the entire Schr\"{o}dinger equation to a
soluble form for a suitably chosen but time-dependent potential. Here we
also address the problem of additional, beyond those that satisfy the linear
Schr\"{o}dinger equation, stationary solutions for the one-dimensional case.
Other solutions can be found when the equations of motion of the
modification are cast into a dynamical system, a stationary point of which
turns out to be a plane wave. The coherent state solution and the Gaussian
wave packet are also identified within this framework. Our findings and
observations are summarized in the last section, where a broader motivation
for this study is also presented.

\section{Main Features of the Modification}

To begin with, let us recall after \cite{Sta1}\thinspace that with the term $%
\gamma (\Delta S)^{2}/2$ implemented the modified action for the
Schr\"{o}dinger equation 
\begin{equation}
S_{mod}({\vec{r}},t)=-\int dt\,d^{3}x\,\left\{ \hbar R^{2}\frac{\partial S}{%
\partial t}+\frac{\hbar ^{2}}{2m}\left[ \left( \vec{\nabla}R\right)
^{2}+R^{2}\left( \vec{\nabla}S\right) ^{2}\right] +R^{2}V+\frac{\gamma }{2}%
\left( \Delta S\right) ^{2}\right\}  \label{1}
\end{equation}
leads to only one new equation 
\begin{equation}
\hbar \frac{\partial R^{2}}{\partial t}+\frac{\hbar ^{2}}{m}\vec{\nabla}%
\cdot \left( R^{2}\vec{\nabla}S\right) -\gamma \Delta \Delta S=0,  \label{2}
\end{equation}
The other equation, that remains unchanged in this Schr\"{o}dinger-Madelung
representation, reads 
\begin{equation}
\frac{\hbar ^{2}}{m}\Delta R-2\hbar R\frac{\partial S}{\partial t}-2RV-\frac{%
\hbar ^{2}}{m}R\left( \vec{\nabla}S\right) ^{2}=0.  \label{3}
\end{equation}
The first of these equations should be viewed as the continuity equation for
the probability density $\rho =R^{2}$ and the current 
\begin{equation}
\vec{j}=\frac{\hbar ^{2}}{m}R^{2}\vec{\nabla}S-\gamma \vec{\nabla}\Delta S.
\label{4}
\end{equation}
The solutions to (2) and (3) are supposed to be normalized in the sense $%
\int d^{3}x\,R^{2}=1$, unless they correspond to the continuous part of the
energy spectrum, in which case the normalization by imposing the periodicity
condition, or the normalization to the delta function is to be applied.
Along with this we need to consider the energy functional which in the
original formulation of Staruszkiewicz \cite{Sta1} was defined as 
\begin{equation}
E_{FT}=\int \,d^{3}x\,\left\{ \frac{\hbar ^{2}}{2m}\left[ \left( \vec{\nabla}%
R\right) ^{2}+R^{2}\left( \vec{\nabla}S\right) ^{2}\right] +\frac{\gamma }{2}%
\left( \Delta S\right) ^{2}\right\} .  \label{5}
\end{equation}
This is a field-theoretical definition of energy. One arrives at it from a
given Lagrangian density $L$, by identifying the energy density of a field
configuration with the time-time component of the canonical energy-momentum
tensor 
\begin{equation}
T_{\nu }^{\mu }=\sum_{i}\left[ \frac{\delta L}{\delta \partial _{\mu
}\varphi _{i}}\partial _{\nu }\varphi _{i}+\frac{\delta L}{\delta \partial
_{\mu }\partial _{\alpha }\varphi _{i}}\partial _{\nu }\partial _{\alpha
}\varphi _{i}-\partial _{\alpha }\left( \frac{\delta L}{\delta \partial
_{\mu }\partial _{\alpha }\varphi _{i}}\partial _{\nu }\varphi _{i}\right)
-\delta _{\nu }^{\mu }L\right] .  \label{6}
\end{equation}
Here $L$ is assumed to depend on a set of fields $\phi _{i}$ and their first
and second order derivatives. The total energy $E_{FT}$ is then given by a
space integral over $T_{0}^{0}$, one example of which is formula (5). The
energy-momentum tensor satisfies the conservation law, 
\begin{equation}
\partial _{\mu }T_{\nu }^{\mu }=0,  \label{7}
\end{equation}
resulting in the energy being a constant of motion unless the parameters of $%
T_{0}^{0}$ (like the potential $V$) depend explicitly on time. In some cases
though even such time-dependence does not prevent $E_{FT}$ from being a
constant of motion.

The Staruszkiewicz modification can be put in the form that involves the
entire wave function $\Psi $ and not only its parts $R$ and $S$. By
observing that $\Delta S=-i\Delta \ln \left( \Psi ^{*}/\Psi \right) /2$, one
can write the Lagrangian density for the modification as 
\begin{equation}
L_{mod}({\vec{r}},t)=\frac{i\hbar }{2}\left( \Psi ^{*}\frac{\partial \Psi }{%
\partial t}-\frac{\partial \Psi ^{*}}{\partial t}\Psi \right) -\frac{{\hbar }%
^{2}}{2m}\vec{\nabla}\Psi ^{*}\vec{\nabla}\Psi -V\Psi ^{*}\Psi +\frac{\gamma 
}{8}\left( \Delta \ln \left( \frac{\Psi ^{*}}{\Psi }\right) \right) ^{2}.
\label{8}
\end{equation}
The factor $\gamma /8$ in the last expression was chosen so as to reproduce
the continuity equation in the form of (2). By altering the density in
question so that the modification term is homogeneous in the wave function $%
\Psi $ similarly as the rest of the Lagrangian one arrives at yet another
modification \cite{Pusz4}. Comparing these two almost identical
modifications can provide a good way to learn about the impact of
nonhomogeneity. From (8) one obtains now the modified Schr\"{o}dinger
equation for $\Psi $ 
\begin{equation}
i\hbar \frac{\partial \Psi }{\partial t}=\left( -\frac{{\hbar }^{2}}{2m}%
\Delta +V\right) \Psi -\frac{\gamma }{8}\left[ \frac{\Delta ^{2}\ln \left( 
\frac{\Psi ^{*}}{\Psi }\right) }{\Psi ^{*}\Psi }\right] \Psi .  \label{9}
\end{equation}
As seen from this equation, the Hamiltonian for the Staruszkiewicz
modification, $H_{SM}$, can also be couched in the hydrodynamic form as 
\begin{equation}
H_{SM}=H_{LSE}+\frac{i\gamma \Delta ^{2}S}{2R^{2}},  \label{10}
\end{equation}
where $H_{LSE}$ is the Hamiltonian for the linear Schr\"{o}dinger equation.
Knowing that in quantum theory the energy of a quantum system is defined as
the expectation value of the Hamiltonian $H$, that is, $E_{QM}=<\Psi |H|\Psi
>=\int d^{3}x\Psi ^{*}H\Psi $, one finds that \footnote{%
The continuity equation implies that for all those situations where 
$R^2$ is normalizable and the gradient of the phase decays fast enough
at infinity, the imaginary term in (11) does not contribute at all by the 
virtue of probability conservation.} 
\begin{equation}
E_{QM}=\int d^{3}x\left\{ \frac{1}{2m}\left[ \left( \vec{\nabla}R\right)
^{2}+R^{2}\left( \vec{\nabla}S\right) ^{2}\right] +VR^{2}+\frac{i\gamma }{2}%
\Delta ^{2}S\right\} .  \label{11}
\end{equation}
Now, even if these two energy functionals are equal for most solutions to
the linear Schr\"{o}dinger equation that are also solutions to this
modification, they do drastically differ for ordinary Gaussian wave packets
for which $\Delta S=g(t)$. Noting that $E_{QM}$ is not the same as $E_{FT}$,
one is faced with a challenge of choosing the right form of the expression
for energy. We will discuss this issue in the next section.

It is reasonable to demand that $E<\infty $, which besides the normalization
condition constitutes another constraint on the solutions to (2) and (3). In
fact, as just pointed out after \cite{Sta1}\thinspace , this constraint
cannot always be satisfied, even for solutions as common to the
Schr\"{o}dinger equation as the Gaussian wave packets if $E_{FT}$ is adopted
for the definition of energy.

To conclude this section, let us observe that without the modification term
equations (2-3) constitute a linear equation, the ordinary Schr\"{o}dinger
equation, in a disguised nonlinear form. One can always transform them back
to the linear equation by employing $S(\Psi ,\Psi ^{*})=-i\ln \left( \Psi
^{*}/\Psi \right) /2$ and $R^{2}(\Psi ,\Psi ^{*})=\Psi \Psi ^{*}.$ On the
other hand, the term $\Delta ^{2}S$ is linear on its own. Therefore,
paraphrasing J. A. Wheeler, one could say that the Staruszkiewicz
modification contains ``nonlinearity without nonlinearity''. It is, in some
sense, a minimal way of introducing nonlinearity and perhaps unique to this
modification in which the linear equation in the disguised nonlinear form is
supplemented by another linear term. The latter however becomes nonlinear in
the Schr\"{o}dinger representation where $\Delta ^{2}S$ turns into $-i\Delta
\ln \left( \Psi ^{*}/\Psi \right) /2$. This incidentally indicates that this
nonlinearity may not be linearizable, i.e., it is not reducible to the
linear Schr\"{o}dinger equation by means of a suitable transformation.

As noted, the discussed modification does not introduce any new dimensional
constants in the system of natural units in three dimensions. In an
arbitrary system of units the dimension of the new coupling constant $\gamma 
$ is the same as that of the product $\hbar c$ or the square of the electric
charge. The coupling constant $\gamma $ can thus be represented as $\gamma =$
$\gamma ^{\prime }\hbar c$, where $\gamma ^{\prime }$ is a dimensionless
parameter in an arbitrary system of units and $c$ is the speed of light. In
a way, if we are to paraphrase J. A. Wheeler once more, this product can be
viewed as introducing the self-interacting electric charge without
introducing any charge at all. However, the combinations $Gm^{2}$ and $\sqrt{%
G\hbar c}m$, where $G$ is Newton's constant, have the same dimension as $%
\hbar c$. This suggests that $\gamma $ could be of the order of either of
the discussed combinations, meaning that $\gamma ^{\prime }$ is of the order
of $1$, although diametrically different values of this parameter cannot be
excluded by these purely dimensional considerations. In the case $\gamma $
is of the order of $Gm^{2}$ or $\sqrt{G\hbar c}m$, one can suspect that the
Staruszkiewicz term $\left( \Delta S\right) ^{2}$ is of a gravitational
origin.

However, the dimensionless nature of the coupling constants of the 
Staruszkiewicz modification is by no means universal for it is only 
characteristic of three dimensions in the natural system of units. 
One can also argue whether the choice of the system of units in which 
$c=1$ is physically justifiable in a modification of the nonrelativistic 
equation and whose nonrelativistic status is sealed by its Galilean 
invariance. We believe that similarly as the original electromagnetic 
underpinnings of the modification cannot be maintained for this would 
violate the $U(1)$-gauge invariance and the Galilean invariance
of the scheme \cite{Pusz2}, so it is more consistent to treat the 
Staruszkiewicz model without referring to any special system of units,
particularly that this is not invariant. Let us note at this point that
there exist quantities that are dimensionless in any system of units. 
Such quantities, to name as an example the fine structure constant, 
can be defined as ratios of other quantities having the same dimensions. 
Yet, the coupling constant in the Staruszkiewicz proposal is not of 
this nature. These shortcomings of the original motivation of 
the Staruszkiewicz proposal can easily be avoided. To this end, we 
suggest that this proposal be reformulated as the simplest nonhomogeneous 
phase modification that derives from a local Galilean invariant Lagrangian. 
The assumption that it is the Lagrangian that should be Galilean invariant 
and not only the equations of motion together with the assumption 
of simplicity leads to a unique modification with a dimensionless 
coupling constant in the natural system of units in three space dimensions. 
It should be noted that in order for the simplicity in question to become 
apparent, the modification needs to be put in the framework of hydrodynamic 
representation of the Schr\"{o}dinger equation. Reformulated in this way, 
the Staruszkiewicz modification can truly stand on its own merits.

\section{Further Analysis of the Modification}

It may be instructive to see how the equations of motion for $R$ and $S$ are
derived for the modification under study. The nontrivial aspect of this
derivation consists in the fact that the action functional (1) from which
these equations are obtained contains derivatives of the second order and
because of this it yields a new boundary term, coming from the $\left(
\Delta S\right) ^{2}$ part of the action. Varying (1) with respect to $R$
and $S$, one arrives at $\delta S_{mod}=I_{1}+I_{2}$ with 
\[
I_{1}=\int_{D}\left[ \left\{ -\hbar \frac{\partial R^{2}}{\partial t}-\frac{%
\hbar ^{2}}{m}\vec{\nabla}\cdot \left( R^{2}\vec{\nabla}S\right) +\gamma
\Delta \Delta S\right\} \delta S+\left\{ \frac{\hbar ^{2}}{m}\Delta R-2\hbar
R\frac{\partial S}{\partial t}-2RV-\frac{\hbar ^{2}}{m}R\left( \vec{\nabla}%
S\right) ^{2}\right\} \delta R\right] , 
\]
\[
I_{2}=\int_{\partial D}\left[ \left( \hbar R^{2}+\frac{\hbar ^{2}}{m}\vec{n}%
\cdot \left( R^{2}\vec{\nabla}S\right) -\gamma \vec{n}\cdot \vec{\nabla}%
\left( \Delta S\right) \right) \delta S+\frac{\hbar ^{2}}{m}\vec{n}\cdot
\left( \vec{\nabla}R\right) \delta R\right] +\int_{\partial D}\gamma \Delta S%
\vec{n}\cdot \delta \left( \vec{\nabla}S\right) , 
\]
where $D$ stands for the domain of integration, being the four-dimensional
space-time, and $\partial D$ for its boundary which is spatial infinity in
the limit $|t|\rightarrow \infty $. The second integral in the last
expression is the new boundary term. To ensure that $\delta S_{mod}=0$ one
needs to assume either that $\delta (\vec{\nabla}S)=0$ on the boundary or
that $\Delta S$ vanishes there.\footnote{%
The other boundary terms in $I_{2}$ are handled in the standard way by
assuming that variations of both $R$ and $S$ vanish at spatial infinity for $%
|t|\rightarrow \infty $.} The latter choice seems to be more attractive, as
one would expect it to promote solutions with finite energy. Unfortunately,
even if $\Delta S$ falls off sufficiently rapidly for large $|\vec{x}|$ or
large $t$, the energy functional (5) or (11) can still have singularity at
the origin. Therefore, in general it seems to be unavoidable to formally
impose $\delta (\vec{\nabla}S)_{|\partial D}=0$, which in fact may
constitute a constraint on the class of acceptable functions for $S$. In any
case, in order to arrive at the Staruszkiewicz modification it is necessary
to ensure that the discussed integral does not contribute to the variation
in question. Of course, if the equations of motion for the modification
are simply postulated and not derived from a Lagrangian density, 
the above analysis does not apply.

The fact that the Staruszkiewicz modification is not homogeneous causes it
to be nonseparable in the sense first discussed by Bia\l ynicki-Birula and
Mycielski \cite{Bial}, also referred to as the weak separability. This
approach is valid only for  noncorrelated subsystems. That this is so was
demonstrated for its extended version in \cite{Pusz1}. The issue of
separability is certainly physically important. The lack of separability
means that even in the absence of interactions the motion of one wave packet
can affect the other in a system consisting of these two packets. This
question has recently been reformulated by Czachor \cite{Czach1} in a novel
``effective'' way which, instead of pure states, uses density matrices as
basic object subjects to a nonlinear quantum evolution. It has been shown
that a much larger class of nonlinearities is allowed than in the Bia\l %
ynicki-Birula and Mycielski approach. In particular, all nonlinearities of
the form $F(|\psi (x)|)$, as for instance the cubic nonlinear
Schr\"{o}dinger equation, are acceptable. As opposed to the weak
separability that assumes that the total wave function of the system is
factorizable to reflect the fact that its subsystems are noncorrelated, the
wave functions in Czachor's approach do not have to obey this condition. For
this reason, the approach in question belongs to the category of strong
separability. As asserted by Czachor \cite{Czach1}, the Staruszkiewicz
modification is separable in this framework.

As noted in the preceding section, the energy definition for the
Staruszkiewicz modification exhibits a curious ambiguity that is absent in
the energy formulation for the linear Schr\"{o}dinger equation. Since quantum
mechanics is a probabilistic theory, it is certainly reasonable to adhere to
the probabilistic interpretation of energy as a quantity that is defined
only in terms of averages. Therefore, the correct form is given by the
expression for $E_{QM}$ which as the mean value of the Hamiltonian
represents the energy according to the standard interpretation of quantum
theory. In keeping with this approach, one dismisses the integral of motion
derived from the conservation of energy-momentum tensor as a viable
candidate for energy in the quantum-mechanical framework, although its
predictive power can still be useful for systems of finite $E_{FT}$. As a
consequence, since $E_{FT}$ is not really an observable in quantum theory
unless it matches the expectation value of the Hamiltonian, one should be
not concerned if this quantity is infinite, even though any infinity in a
physical theory is rather disturbing. A less radical alternative approach
would consist in reconciling these two different objects.

Let us now argue that compared to $E_{FT}$, $E_{QM}$ seems to be more
acceptable on the grounds of physical consistency. In the Staruszkiewicz
modification the phase is decoupled from the amplitude of the wave function
for, as seen from the equations of motion, it remains a physically
determined quantity, having a life of its own even if $R=0$! In linear
quantum mechanics, except for the nodal points that constitute a discrete
set of points, a situation like that does not take place. In fact, it seems
to be meaningless to talk about the phase when the amplitude vanishes in a
finite region of space and so do the equations of motions leaving no room to
determine the evolution in this region. However, in the Staruszkiewicz
modification the phase can, in principle, be determined through the
equations of motion even if it is not accompanied by the amplitude. To see
what consequences this can possibly entail let us consider a one dimensional
wave packet moving freely in space until it encounters a totally
impenetrable wall, an infinitely large potential barrier. Now, according to
the standard interpretation of quantum mechanics, the amplitude of the
packet has to vanish beyond the wall, but this does not apply to the phase
which can be nonzero there. In linear quantum mechanics, the energy
associated with the phase beyond such an infinite wall is zero as long as
its gradient is finite which is to be expected as the probability current $%
\vec{j}=R^{2}\vec{\nabla}S$ should vanish there, too. This is not so in the
Staruszkiewicz modification unless one requires that the energy contribution
due to the nonlinear part is zero in the region beyond the wall. The
vanishing of the probability current does not entail this if one identifies
the energy with $E_{FT}$. Therefore, one is faced with a transmission of
energy through an infinitely large barrier that may not be accompanied by a
flow of the probability current. It is clear that this situation is not
physically sound, but fortunately it can be amended if one defines energy as 
$E_{QM}$, which seems to be the first argument in favor of this definition
of energy. Now, assuming that the current is a continuous function and
vanishes on the wall implies that it vanishes also beyond it as $\Delta
^{2}S $ is zero everywhere beyond the wall by the virtue of the equations of
motion or just the continuity equation. In other words, $\Delta ^{2}S=0$
implies that the current is constant beyond the wall, but owing to its
continuity it can only be zero there. Moreover, the energy associated with
the phase beyond the wall is zero as is the energy of a ``quantum mechanical
system'' in the entirely free space in the absence of the amplitude if
identified with $E_{QM}$. Unless the phase alone is really proven to be
endowed with energy, this observation once again supports $E_{QM}$ as a more
physically reasonable expression for the energy of a quantum-mechanical
system. Still, even if no energy can be associated with the phase in the
limit of vanishing amplitude, it is not out of the question that it can be
detected by some diffraction or interference phenomena similar to the
Aharonov-Bohm effect \cite{Ahar}. To summarize our reasoning, electing $%
E_{QM}$ over $E_{FT}$ stems from a very simple yet physically respectable 
requirement that no information, energy in particular, is allowed to be
transmitted through an infinitely large potential barrier.

The energy functional $E_{QM}$ contains an imaginary component. Since the
energy is supposed to be a real quantity one might want to require that this
part does not contribute to the total energy, which imposes a constraint on
physically acceptable states in this particular nonlinear model of quantum
mechanics. That this constraint is not necessarily very restrictive has
already been demonstrated for the Gaussian wave packets. Since for such
packets $E_{FT}$ is infinite, they would have to be excluded from legitimate
solutions to the modification if this energy definition were employed. Yet,
they are perfectly fine on the energetic grounds if one uses the
quantum-mechanical definition of energy. Moreover, the continuity equation
coupled with the assumption of probability conservation implies the
vanishing of this term for a large class of physically plausible situations.
The proposed condition of real energy is equivalent to the selection of
observables in nonlinear quantum mechanics for which contributions of
nonlinear parts vanish on normalized states \cite{Czach2}. On the other
hand, complex energy is not in the least foreign to linear quantum mechanics
and therefore it is certainly justifiable to entertain the possibility of
retaining it in the modification under discussion. In linear quantum
mechanics, complex energy can appear as a result of implementing complex or
``optical'' potentials usually invoked to phenomenologically model
absorption in scatttering processes. The complex potentials have also been
used to describe decoherence \cite{Mens}. Since the Staruszkiewicz
modification introduces complex potential-like terms in a natural manner,
one finds it tempting to allow states with complex energy if only in the
hope that they can mediate in the process of decoherence.Such states,
however, would have to be rather exotic because, as noted, in the majority
of physically reasonable situations one expects the integral over the
imaginary term in question to vanish.

As we see, the properties of observables depend on the space of states in
which they are defined, which also applies to linear quantum mechanics,
where, for instance, the self-adjointness of an operator depends on its
domain. In line with this approach, one can attempt to reconcile $E_{QM}$
with $E_{FT}$ by choosing a domain in which they are equal for each function
in the domain. This alternative seems to be coming as close as possible to
the original Staruszkiewicz formulation by retaining $E_{FT}$ as a
physically relevant object, but it is probably more restrictive than the
formulation just proposed for it requires not only that $E_{FT}$ be finite
and $E_{QM}$ be real, but, in addition, that these two quantities be equal.
In what follows, for the sake of exploring all possibilities, we will
consider both of these formulations. In fact, none of them seems to be
decidedly more convincing than the other. We will refer to the alternative
in question as maximal to contrast it with a moderate formulation that is
concerned only with the expectation value of a Hamiltonian treating $E_{FT}$
as useful but not a crucial quantity in the quantum-mechanical scheme it
adheres to. As we will see, all the nontrivial solutions that we have found
belong to the moderate formulation, although one of them is shared by the
original formulation as well. This lack of uniqueness in the definition of
energy clearly marks the difference between linear quantum mechanics and its
nonlinear models where uniqueness seems to be an exception rather than a
rule.

It is possible to adopt yet another approach that takes care of the weak
nonseparability problem and the problem of uniqueness of energy at the same
time. This is accomplished by considering as legitimate only those solutions
to the modified Schr\"{o}dinger equation that satisfy the equation $\Delta
^{2}S=0$. Even if the strong separability seems to offer a more
all-encompassing approach, it is the weak separability that is a more
stringent condition and for this reason one can use it as a criterion for
the formulation discussed. We will call this formulation minimal for the
domain of its solutions is expected to be the smallest of all the
formulations considered. It overlaps with that of the maximal formulation in
the case $\Delta S=0$ and is entirely contained within the domain of the
moderate formulation. Despite its name, this approach ensures the maximum of
physically desirable properties, including, as we will soon see, the
standard classical limit of quantum theory.

Let us now comment on the semiclassical approximation for the modification.
Formally, one performs this approximation by rescaling $S$ to $S/\hbar $,
followed by expanding $S$ in a series $S=S_{0}+\hbar S_{1}+\hbar ^{2}S_{2}+$%
...One obtains in this way an expansion of the Schr\"{o}dinger equation in
the powers of the Planck constant. The lowest order approximation term to
the continuity equation of the linear Schr\"{o}dinger equation is
proportional to $\hbar $. This, however, is not the case for its modified
counterpart whose expansion in the powers of $\hbar $ starts with $\hbar
^{-1}$. The next leading term is proportional to $\hbar ^{0}$. As a result,
we obtain that $\Delta ^{2}S_{0}=\Delta ^{2}S_{1}=0$. In practice, this
means that, in a very good first approximation, the solutions to (2) can be
sought for among wave functions whose phase $S$ satisfies the condition $%
\Delta ^{2}S=0$. One could hope to find in this way even exact solutions to
(2) and (3), with, perhaps, some additional but physically well justified
potential $V$. As we will see in the next section, this hope turns out to be
justified.

The classical limit of this modification in the sense of the Ehrenfest
theorem may not always exist since the standard Ehrenfest theorem of linear
quantum mechanics is altered by nonlinear corrections. As worked out in \cite
{Pusz1}, the general form of the modified Ehrenfest relations is 
\begin{equation}
m\frac{d}{dt}\left\langle \vec{r}\right\rangle =\left\langle \vec{p}%
\right\rangle +F_{1},  \label{12}
\end{equation}
\begin{equation}
\frac{d}{dt}\left\langle \vec{p}\right\rangle =-\left\langle \vec{\nabla}%
V\right\rangle +F_{2},  \label{13}
\end{equation}
where 
\begin{equation}
F_{1}=\frac{2m}{\hbar }\int \,d^{3}x\vec{r}R^{2}H_{I},  \label{14}
\end{equation}
\begin{equation}
F_{2}=\int \,d^{3}xR^{2}\left( 2H_{I}\vec{\nabla}S-\vec{\nabla}H_{R}\right) .
\label{15}
\end{equation}
$H_{R}$ and $H_{I}$ represent the real and imaginary parts of the nonlinear
part of the Hamiltonian of the modified Schr\"{o}dinger equation,
respectively. In the derivation of the last formula it was assumed that $%
\int d^{3}x\vec{\nabla}\left( R^{2}H_{I}\right) =0$ and because of that this
term was discarded. Since $H_{I}=\gamma \Delta ^{2}S/2R^{2}$, this is
tantamount to assuming that $|\vec{x}|^{2}\Delta ^{2}S$ vanishes at
infinity. This, in turn, is implied by the condition that the energy $E_{QM}$
be finite, and it is satisfied for all those configurations that are
square-integrable as can be seen from the continuity equation given the
probability is conserved. Had $\int d^{3}x\vec{\nabla}\left(
R^{2}H_{I}\right) $ been nonzero, it would introduce an imaginary component
to $F_{2}$. Now, since $H_{R}=0$ and $H_{I}=\gamma \Delta ^{2}S/2R^{2}$, the
Ehrenfest relations for the modification in question read 
\begin{equation}
m\frac{d}{dt}\left\langle \vec{r}\right\rangle =\left\langle \vec{p}%
\right\rangle +\frac{\gamma m}{\hbar }\int \,d^{3}x\vec{r}\Delta ^{2}S,
\label{16}
\end{equation}
\begin{equation}
\frac{d}{dt}\left\langle \vec{p}\right\rangle =-\left\langle \vec{\nabla}%
V\right\rangle +\gamma \int \,d^{3}x\Delta ^{2}S\vec{\nabla}S.  \label{17}
\end{equation}
These relations are Galilean covariant and their nonlinear contributions are
also Galilean invariant for the wave functions that are normalizable and do
not violate the probability conservation.

We see that, in general, the nonlinear corrections to the Ehrenfest
relations do not vanish, leading to a different classical limit than in
linear theory. This feature of the modification is shared by other nonlinear
generalizations of the Schr\"{o}dinger equation, the only notable exception
from this rule being the Bia\l ynicki-Birula and Mycielski modification. It
suggests that the Staruszkiewicz modification is not linearizable and as
such may describe phenomena that cannot be captured by the linear
Schr\"{o}dinger equation. On the other hand, perhaps due to its simplicity,
the relevance of the Ehrenfest theorem tends to be overestimated. As argued
in \cite{Ball}, this theorem is neither sufficient nor necessary to
characterize the classical regime of quantum theory.

However, one can also entertain the possibility that the classical limit of
quantum world does not always exist in the form suggested by classical
theory. This might offer an explanation of some anomalous or otherwise
unexplained phenomena some of which owing to their strangeness remain
outside the realm of accepted exact science. As a matter of fact, since even
the quantization procedure is sometimes ambiguous, a unique classical limit
of quantum theory is something that is enforced by rather than naturally
emerges from it. What is remarkable in the modification discussed, it is
that the departure from the standard classical limit is very subtle and, at
the same time, pretty elusive due to the very elusive nature of the phase.
The departure in question, unlike, for instance, in some other modification
that possesses the same property \cite{Pusz4}, is also entirely defined by
the phase.\footnote{%
Other modifications, as for instance the Doebner-Goldin \cite{Doeb}
modification, can also have a non-standard classical limit. However, since
the Doebner-Goldin modification describes only irreversible and dissipative
quantum systems it cannot be treated on the same footing as the
Staruszkiewicz modification. Moreover, it is also less unique for it
involves as many as $6$ dimensional parameters.}

Finally, let us make two more comments utilizing yet another rather
straightforward modification of the Schr\"{o}dinger equation, a
relativistically modified Schr\"{o}dinger equation which follows when the
expression for the energy of a single particle in terms of its momentum is
taken in the leading relativistic approximation 
\begin{equation}
E=\frac{p^{2}}{2m}-\frac{p^{4}}{8m^{3}c^{2}},  \label{18}
\end{equation}
and the first quantization is performed. The Schr\"{o}dinger equation for
this modification, 
\begin{equation}
i\hbar \frac{\partial \Psi }{\partial t}=\left( -\frac{\hbar ^{2}}{2m}\Delta
-\frac{\hbar ^{4}}{8m^{3}c^{2}}\Delta ^{2}\right) \Psi =0,  \label{19}
\end{equation}
when put in the Madelung representation involves additional contributions to
the continuity equation in both $S$ and $R$. It is one of the reasons that
this particular modification can be considered more general than the
Staruszkiewicz modification. However, despite its generality, the
relativistic modification does not contain Staruszkiewicz's modification as
a special case. This can be seen by comparing the dimensions of the coupling
constants of these modifications. Moreover, unlike the Staruszkiewicz
modification, the relativistic modification is homogeneous of degree one in
the wave function and is linear. The other comment consists in noting that
the Gaussian wave packets, which are not solutions to the Staruszkiewicz
modification for energetic reasons\footnote{%
As found in \cite{Sta1}, due to a slow convergence of their phase in the
infinity, the energy $E_{FT}$ is not finite for the Gaussian wave packets.},
are also not solutions to (19), which indicates that it is rather difficult
to preserve wave packet solutions in other extensions of this equation as
well. Therefore, the modification under discussion in its original
formulation should not necessarily be viewed as pathological in this
respect. However, what is really unacceptable here is that $E_{FT}\neq E_{QM}
$. This causes the wave packets to be excluded from the maximal formulation
of the Staruszkiewicz modification.

\section{Solutions}

It is not easy to find solutions to the modification discussed here. The
main obstacle to this stems from the nonhomogeneity of (2). Even if such
solutions are found they may not necessarily yield to a clear physical
interpretation. Therefore, it should not come as a surprise that the
physically best understood solution we discovered so far represents a case
when the nonhomogeneity does not interfere.

As suggested in the previous section, one should look for the solutions to
the Staruszkiewicz modification among the phenomena for which the square of
the Laplacian of the phase $S$ vanishes. We do not intend to present here a
complete solution to the equation $\Delta ^{2}S=0$ together with a
discussion of its applicability to this modification, reserving it for
another publication \cite{Pusz5}.\thinspace Instead, we prefer to show the
simplest possible and physically important solution to the equation $\Delta
S=0$ that, obviously, satisfies the previous one and ensures the finite
unique energy in each of the discussed formulations of this modification.
Such a solution is provided by a coherent state \cite{Schr}. It is easy to
check that indeed the coherent state described by 
\begin{equation}
R^{2}=\frac{1}{\sqrt{\pi }x_{0}}\exp \left[ {-\frac{\left( x-x_{0}\sqrt{2}%
\cos (\omega t-\delta )\right) ^{2}}{{x_{0}}^{2}}}\right]  \label{20}
\end{equation}
and 
\begin{equation}
S=-\left( \frac{\omega t}{2}-\frac{|\alpha |^{2}}{2}\sin 2\left( \omega
t-\delta \right) +\frac{\sqrt{2}|\alpha |x}{x_{0}}\sin \left( \omega
t-\delta \right) \right)  \label{21}
\end{equation}
represents a solution to equations (2) and (3) in one dimension in the
potential of a simple harmonic oscillator $V=m\omega ^{2}x^{2}/2$. Here $%
x_{0}=(m\omega )^{-1/2}$, while $\alpha $ and $\delta $ are arbitrary
constants, complex and real, respectively. We also put $\hbar =1$. The three
dimensional case can be worked out as a product of one-dimensional
amplitudes with $\omega $ and the phase $\delta $ appropriately redefined.

Another physically interesting solution to (2) in one dimension can be found
for a general negative $\gamma =-|\gamma |$. In this case, (2) becomes 
\begin{equation}
\hbar \frac{\partial R^{2}}{\partial t}+\frac{\hbar ^{2}}{m}\left(
R^{2}S_{x}\right) \,_{x}+|\gamma |S_{xxxx}=0.  \label{22}
\end{equation}
We will now show that this equation can be reduced to one of the standard
forms of the celebrated Korteweg-de Vries equation 
\begin{equation}
u_{t}+3(u^{2})_{x}+u_{xxx}=0.  \label{23}
\end{equation}
To this end, let us make identifications $S_{x}=R^{2}=du(x)$, where $d$ is a
free constant whose dimension is to be worked out. Because the dimensions of 
$S_{x}$ and $R^{2}$ are equal, the dimension of $t$ has to be meter$^{3}$.
This makes it impossible to maintain the natural system of units if we
insist that $\gamma $ be dimensionless to preserve the fundamental idea of
the modification in its original formulation. We can assume that either $%
\hbar =1$ or $c=1$ but not both. Such a compromise is unavoidable in one
dimension. We will assume that $\hbar =1$ which implies that the speed of
light has the dimension of meter$^{-2}$ and the dimension of $m$ has to be
that of meter. Now, it is certainly in the spirit of this modification to
make the coefficient of $(u^{2})_{x}$ dimensionless and, in fact, it has to
be so for (22) to reduce to the form of (23). Therefore, $d/m|\gamma |=3$
which does not impose any constraints on $|\gamma |$ because of the freedom
of choosing $d$, but fixes the dimension of $d$ to be that of meter.
Moreover, $t$ gets rescaled to $t^{\prime }=|\gamma |t$. The most physically
attractive solution to the Korteweg-de Vries equation is known to be a
soliton of the form 
\begin{equation}
u_{sol}(\theta ;k)=\frac{k^{2}}{2\cosh ^{2}\frac{\theta }{2}},  \label{24}
\end{equation}
where $\theta =k\xi +\tau $, $\xi =x-|\gamma |k^{2}t$, and $\tau $ is an
arbitrary phase. The dimension of $k^{2}$ is the same as the dimension of $%
u(x)$, meter$^{-2}$. This solution is normalizable and the normalization
condition, $\int dxR^{2}(x)=d\int dxu(x)=1$, implies that $dk=1/2$.
Consequently, since $d/m|\gamma |=3$, the ``speed'' of the soliton is $%
v_{s}=|\gamma |/4d^{2}=1/36|\gamma |m^{2}$. From (3), we find out that the
entire Schr\"{o}dinger-Madelung system of equations can support the KdV
soliton only in the presence of the potential 
\begin{equation}
V(\theta ;k)=\frac{1}{8m\cosh ^{4}\frac{\theta }{2}}\left[ \left(
k^{2}+4Em\right) \cosh ^{4}\frac{\theta }{2}+2k^{2}\left( 2md|\gamma
|k^{2}-1\right) \cosh ^{2}\frac{\theta }{2}-d^{2}k^{4}\right] ,  \label{25}
\end{equation}
where $E$ is a constant of the same dimension as energy, meter$^{-3}$. In
this potential $E_{FT}$ is finite (and also constant!), while $E_{QM}$ is
real and constant as well, however they are not equal. It is not clear how
this potential can be physically realized. The solution under consideration,
similarly as the Gaussian and coherent state wave packets, does not alter
the standard Ehrenfest relations.

The solutions discussed above were non-stationary. As observed in \cite
{Pusz1}, the Staruszkiewicz modification, as opposed to its extension \cite
{Pusz1} and some other notable nonlinear modifications of the
Schr\"{o}dinger equation \cite{Bial, Doeb}, does not affect
stationary states of quantum mechanical systems determined by the
Schr\"{o}dinger equation. This means that wave functions describing these
states are also solutions to the modified equation with unchanged energy
levels. It is this particular property that makes the Staruszkiewicz
modification stand out. As a function of time and position, the phase of
these states is given by $S=-Et/\hbar +\sigma (\vec{x})$, but its position
dependence encoded in $\sigma (\vec{x})$ is physically inconsequential.
Usually, $\sigma (\vec{x})$ is at most a linear function of spatial
coordinates. However, it is conceivable that there exist other stationary
states for which the phase is also a physically nontrivial function of
position in the sense that, for instance, the phase affects the energy of
the system. We will now show how one can approach the problem of finding
such solutions in the one-dimensional case.

To find stationary solutions to the Staruszkiewicz modification one starts
with the condition 
\begin{equation}
\frac{\partial R^{2}}{\partial t}=0.  \label{26}
\end{equation}
This condition when applied to (2) and (3) together with the assumption of
time-independent potential implies that $\partial S/\partial
t=const=-E/\hbar $, where $E$ is to be identified with the energy of a
quantum-mechanical system. For one-dimensional stationary problems, assuming 
$\hbar =1$, equations (2) and (3) reduce to 
\begin{equation}
u^{\prime \prime }+aR^{2}u+c=0,  \label{27}
\end{equation}
\begin{equation}
R^{\prime \prime }+2m\left( E-V-bu^{2}\right) R=0,  \label{28}
\end{equation}
where $u=S_{x}$, $a=-1/\gamma m$, $b=1/2m$, and $c$ is a free constant that
can be zero.

One can seek solutions to this system in two different ways. One is by
selecting a normalizable $R$ representing a solution to the linear
Schr\"{o}dinger equation in some potential $V$. Treating this as an Ansatz
one attempts to solve the first of the above equations for $u$. This method
will usually lead to a complicated equation for $u$ that may not be solvable
in an analytical manner even for a simple $R$. Once $u$ is determined, the
elected $R$ can be interpreted as the solution to the linear Schr\"{o}dinger
equation, but in the effective potential $V_{eff}=V-bu^{2}$. The other way
boils down to picking $u$ that ensures that (27) leads to a normalizable and
positive $R^{2}$. Inserting now $R$ and $u$ into (28) gives us the potential
that supports this configuration. Nevertheless, the solutions found in these
ways may not necessarily yield to a simple physical interpretation. For
instance, by employing the second method for $c=0$, $a=-2$, and $%
u(x)=1+x^{2} $ one obtains $R^{2}=1/(1+x^{2})$. This configuration is
obviously well localized and thus normalizable. It exists in the effective
potential $V_{eff}=\left( 2x^{2}-1\right) /2(x^{2}+1)^{2}-\gamma \left(
1+x^{2}\right) ^{2}$ which is unbounded from below as $\gamma $ is positive.
The energy $E_{QM}$ of this stationary solution is finite and equal zero
even if the potential $V_{eff}$ can be negative in the whole domain for a
sufficiently large $\gamma $. However, its field-theoretical energy is
infinite. The configuration in question does not affect the standard
Ehrenfest relations.

Similar solutions appear in the linear equation as well. For example, one
can obtain exactly the same solution as considered above in the linear case
if $c\neq 0$. In the modification discussed, the solutions in question
might, however, play more important and perhaps a novel role due to a more
pronounced contribution of the phase to the equations of motion. The major
problem with these solutions in both linear and nonlinear theory is that
they seem to be vulnerable to generic small perturbations. The perturbed
configurations can have infinite energy and therefore cannot be considered
physical. This implies that the functional domain of stability of the
solutions discussed is very small. In fact, they may not be stable at all,
that is, an infinitely small perturbation can destroy them. Such a statement
would require a formal proof, but since we cannot provide it here, we offer
it as a conjecture.\footnote{%
Since the solutions in question appear in both linear and nonlinear theory,
we will discuss them in greater detail in a separate paper hoping to resolve
this issue in it \cite{Pusz6}.} Consequently, it is safe to say that no
other normalizable stationary solutions of physical interest beyond those to
the linear equation appear in the Staruszkiewicz modification. In principle,
one can also address the issue under consideration as a spectral problem of
equation (9). However, in the case of nonlinear equations, each spectral
problem requires a separate mathematical analysis as, unlike for linear
operators on Hilbert spaces, there is not any general theory that could
tackle such problems.

One of the most general frameworks in which to seek solutions to the
modified Schr\"{o}dinger equation is that of dynamical systems. For
simplicity, we will employ it here only for the one-dimensional case, in
which one has 
\begin{equation}
\omega ^{\prime }=\frac{1}{\gamma }\left[ \frac{\partial \epsilon }{\partial
t}+\frac{1}{m}\epsilon \left( 2\zeta \eta +\nu \right) \right] ,  \label{29}
\end{equation}
\begin{equation}
\zeta ^{\prime }=2m\left( \frac{\partial S}{\partial t}+V(x,t)+\frac{1}{2m}%
\eta ^{2}\right) -\zeta ^{2},  \label{30}
\end{equation}
\begin{equation}
\epsilon ^{\prime }=2\epsilon \zeta ,  \label{31}
\end{equation}
\begin{equation}
\eta ^{\prime }=\nu ,  \label{32}
\end{equation}
\begin{equation}
\nu ^{\prime }=\omega ,  \label{33}
\end{equation}
with the variables $\eta $, $\epsilon $, $\zeta $, $\nu $, and $\omega $
defined as $\eta =S^{\prime }=dS/dx$, $\epsilon =R^{2}$, $\zeta =R^{\prime
}/R$, $\nu =S^{\prime \prime }$, and $\omega =S^{\prime \prime \prime }$. It
should be noted that the dynamical system (29-33) ``evolves'' in $x$, the
actual time variable $t$ being a parameter.

Looking for stationary points of this system, we dismiss $\epsilon =0$ as
unphysical, which leaves us with an alternative $\zeta =\omega =\nu =0$ and $%
\partial \epsilon /\partial t=0$, and $\partial S/\partial t+V(x,t)+\eta
^{2}/2m=0$. As can easily be verified, the plane wave described by $S=-Et+kx$
and $R=1$, with the normalization condition realized by imposing the
periodic boundary conditions, is a stationary point of this dynamical system.%
\footnote{%
This observation was first made by A. Z. G\'{o}rski and P. O. Mazur in 1984 
\cite{Maz}\,.} In fact, it is probably the only such a point.

As a way to demonstrate the capacity of this method let us now find out how
the other solutions to the nonlinear Schr\"{o}dinger equation discussed in
the present paper can be identified within this framework. The coherent wave
packet with $R^{2}=\epsilon $ and $S$ given by (20-21) corresponds to $\eta
=-\sqrt{2}|\alpha |\sin (\omega t-\delta )/x_{0}$, $\zeta ={-\left( x-x_{0}%
\sqrt{2}\cos (\omega t-\delta )\right) /{x_{0}^{2}}}$, and $\nu =\omega =0$.
One deals here with an effectively two-dimensional phase space (for any
fixed $t$) spanned by $\zeta $ and $\epsilon $ as they are the only
variables allowed to change in ``time'' $x$. For comparison, the Gaussian
wave packet with $R$ and $S$ specified by 
\begin{equation}
R=\left[ \frac{mt_{0}}{\pi \left( t^{2}+t_{0}^{2}\right) }\right] ^{1/4}\exp
\left[ -\frac{mt_{0}x^{2}}{2\left( t^{2}+t_{0}^{2}\right) }\right] 
\label{34}
\end{equation}
and 
\begin{equation}
S=\frac{mtx^{2}}{2\left( t^{2}+t_{0}^{2}\right) }-\frac{1}{2}\arctan \frac{t%
}{t_{0}},  \label{35}
\end{equation}
lives in an effectively three-dimensional phase space. Indeed, in this case
the motion takes place in variables $\eta =mt|x|/\left(
t^{2}+t_{0}^{2}\right) $, $\zeta =-mt_{0}|x|/\left( t^{2}+t_{0}^{2}\right) $%
, and $\epsilon $, the other variables being $\nu =mt/\left(
t^{2}+t_{0}^{2}\right) $ and $\omega =0$. They both are constants for the
actual time fixed. With the time allowed to run, the manifolds on which the
evolutions of these packets occur are three- and four-dimensional,
correspondingly. This comparison of ordinary wave packets with the coherent
states is just another way of demonstrating that the phase space of the
latter is smaller than that of the former. It is due to this very fact that
the coherent wave packets are also called ``the minimal phase space
packets''. The KdV soliton does not represent any particular plane in this
phase space.

\section{Conclusions}

The paper presented was aimed at recalling and a further discussion of the
nonlinear modification of the Schr\"{o}dinger equation proposed by
Staruszkiewicz. Historically motivated by a longitudinal modification of the
electromagnetic action, here it has been treated as postulated on the
grounds of physical consistency and certain attractiveness similarly as
other notable nonlinear modifications of this equation \cite{Bial, Doeb}. 
Unlike these, the Staruszkiewicz modification does not introduce any
new dimensional constants in the system of natural units as long as it is
formulated in three dimensions. As argued, due to the energy ambiguity
typical of nonhomogeneous modifications of the Schr\"{o}dinger equation,
depending on which physically relevant properties one decides to maintain,
one can come up with three different formulations of the Staruszkiewicz
modification, although we believe that the arguments presented favor the
moderate formulation which assumes that the correct definition of energy is
given by the expectation value of the Hamiltonian and rejects any other
definitions as spurious. It is within this formulation that the largest
number of solutions can be found, including the Gaussian wave packet that
could not be incorporated in the original version of the modification. The
formulation in question admits also some other physically interesting
solutions, among them the coherent state for the potential of harmonic
oscillator, shared though by all the formulations, and the KdV soliton for
another suitably chosen but time-dependent potential. None of these
solutions alters the standard Ehrenfest relations.

One can speculate that Staruszkiewicz's modification describes some part of
reality that due to rather an elusive nature of the phase of wave function
has escaped our realization. In fact, until relatively recently the phase
had been treated as an object of rather secondary importance to the
understanding of quantum-mechanical systems. However, it turned out that in
a number of physical situations such a position could not be maintained for
it would lead to an incomplete physical description. The phase of the wave
function encodes relevant information on the quantum evolution as manifested
in the Aharonov-Bohm effect \cite{Ahar}\thinspace and other phenomena of
similar nature \cite{Berr, Anan1} (see also \cite{Anan2}\thinspace for a
more complete list of references). Another motivation to study this
modification stems from the long-standing problem of the collapse of wave
function. Since it is rather reasonable to expect that the phase plays an
important part in this process, one hopes that the modification discussed
can offer some insight into the physics of this phenomenon.

In all its formulations, the Staruszkiewicz modification introduces very
minimal changes to linear quantum theory. Not only are the stationary states
and thus also the atomic structure unchanged, but also the energy of
nonstationary phenomena remains the same in most if not all physically
relevant cases. Despite these, the modification in question differs from
linear quantum mechanics in its classical limit. It is, however, not out of
the question that to solve the riddle of the measurement problem and perhaps
to understand other phenomena currently viewed as anomalous, the idea of
modifying the classical limit of quantum theory must be seriously
considered. Being a fundamental theory of nature, it is the quantum theory
that should tell us what the world is like in the classical limit. Imposing
such a limit on the basis of our apparently ``classical'' collective
experience may deprive us of the understanding of phenomena that are not so
common to this experience and thus, sometimes, not even widely regarded as
real. As noted by Einstein, it is theory that is supposed to tell us what is
observable. The Staruszkiewicz modification offers the subtlest conceivable
proposal to extend the classical limit. Moreover, as partially indicated by
this non-standard limit, it is very plausible that this modification is not
linearizable and therefore it could describe phenomena that cannot be captured
by the linear theory.

\section*{Acknowledgments}

I am very grateful to Professor Andrzej Staruszkiewicz for a discussion 
of his modification and to Dr. Marek Czachor for a useful and stimulating
exchange of email correspondence. I would also like to thank Professor 
Pawe{\l } O. Mazur for bringing to my attention the work of Professor 
Staruszkiewicz discussed here. This work was partially supported by 
the NSF grant No. 13020 F167 and the ONR grant R\&T No. 3124141.

\end{document}